\documentclass[%
 reprint,
 amsmath,amssymb,
 aps,
 pre,
 showpacs
]{revtex4-2}

\usepackage{graphicx}
\usepackage{dcolumn}
\usepackage{bm}

\usepackage[utf8]{inputenc}

\begin{document}

\title{Interplay between tie strength and neighbourhood topology in complex networks: Granovetter’s theory and beyond}

\author{Maciej J. Mrowinski$^*$}
\author{Kamil P. Orzechowski}
\author{Agata Fronczak}
\author{Piotr Fronczak}
\affiliation{%
    Warsaw University of Technology, Faculty of Physics,\\
    ul.~Koszykowa 75, 00-662 Warsaw, Poland\\
    $^{*}$ Corresponding author; email: maciej.mrowinski@pw.edu.pl
}%

\begin{abstract}
Granovetter's weak ties theory is a very important sociological theory according to which a correlation between edge weight and the network's topology should exist. More specifically, the neighbourhood overlap of two nodes connected by an edge should be positively correlated with edge weight (tie strength). However, some real social networks exhibit a negative correlation - the most prominent example is the scientific collaboration network, for which overlap decreases with edge weight. It has been demonstrated that the aforementioned inconsistency with Granovetter's theory can be alleviated in the scientific collaboration network through the use of asymmetric measures. In this paper, we explain that while asymmetric measures are often necessary to describe complex networks and to confirm Granovetter's theory, their interpretation is not simple, and there are pitfalls that one must be wary of. The definitions of asymmetric weights and overlaps introduce structural correlations that must be filtered out. We show that correlation profiles can be used to overcome this problem. Using this technique, not only do we confirm Granovetter's theory in various real and artificial social networks, but we also show that Granovetter-like weight-topology correlations are present in other complex networks (e.g. metabolic and neural networks). Our results suggest that Granovetter's theory is a sociological manifestation of more general principles governing various types of complex networks.
\end{abstract}
\pacs{89.75.Hc, 89.65.-s, 89.75.Fb, 89.70.+c}

\maketitle

\section{Introduction}

While this is not always the case, the weights of edges in networks are usually quantitative expressions of the mutual relationship between nodes. Be it the number of scientific collaborations between authors or the number of mentions in a social network, the weight of an edge often signifies the strength of the connection between nodes. It stands to reason that this strength must, in some way, correlate with the network's structure - specifically, with the relative position of nodes and their neighbourhoods within the network.

\begin{figure}[t]
    \centering
    \includegraphics[width=\linewidth]{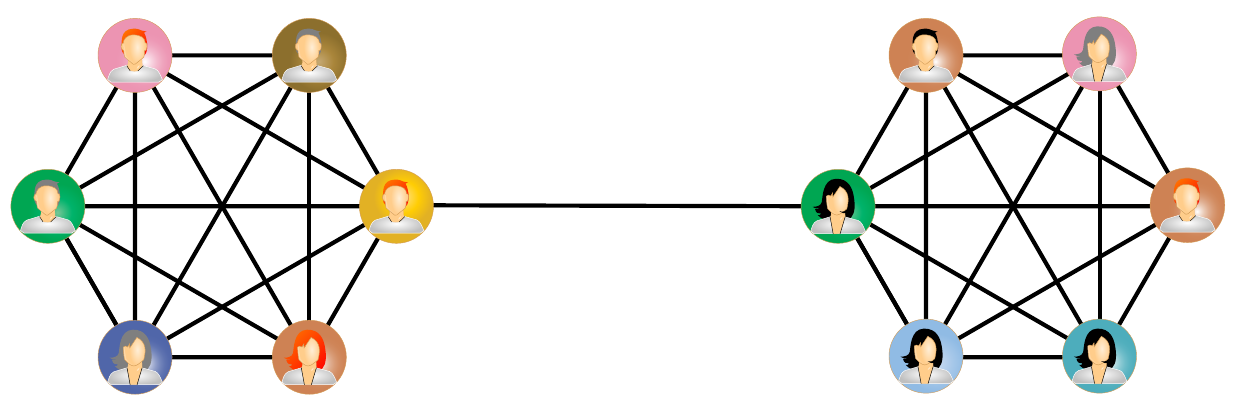}
    \caption{An example illustrating Granovetter's theory. A detailed description can be found in the text.}\label{fig:granovetter}   
\end{figure}
Mark Granovetter, in his famous work Strength of Weak Ties \cite{granovetter1973strength, granovetter2018getting}, introduced a theory which aims to explain the aforementioned link between the weights of edges and the topology of the network. An example that illustrates Granovetter's hypothesis can be found in Fig.~\ref{fig:granovetter}, which shows two fully connected clusters of nodes. According to Granovetter, since virtually all nodes in each cluster have the same neighbourhoods, we should expect that edge weights (tie strengths) within clusters will be high. The clusters are also connected by a single edge. Edge weight of this connection should be low, as nodes at both sides of the link do not share any neighbours. Granovetter's theory also states that weak ties, like the one connecting the clusters in our example, are crucial to the diffusion of information in the network and nodes that have access to such ties have an advantage over those that do not. In this work, however, we are only interested in the first part of the theory, that is in weight-topology correlations.

In more formal terms, the first part of Granovetter's theory states that edge weight is positively correlated with the overlap of the neighbourhood of two connected nodes. The overlap between neighbourhoods of node $i$ and node $j$ is defined in the following way \cite{Onnela2007structure}
\begin{equation}
O_{ij} = \frac{n_{ij}}{(k_i - 1) + (k_j - 1) - n_{ij}},
\end{equation}
where $n_{ij}$ is the number of common neighbours of nodes $i$ and $j$, $k_i$ and $k_j$ are degrees of nodes $i$ and $j$. It is worth noting that overlap, as defined above, is a symmetric measure
\begin{equation}
O_{ij} = O_{ji},
\end{equation}
and we will refer to it as symmetric overlap to emphasize this fact. Similarly, weights $w$ are also assumed to be symmetric, that is
\begin{equation}
w_{ij} = w_{ji}.
\end{equation}

Granovetter's theory in this form - that is, a monotonically increasing relation between $O_{ij}$ and $w_{ij}$ -  has been empirically confirmed, to various extents, in real social networks \cite{Easley2010, Onnela2007structure, Eagle2010network, Pajevic2012organization, Grabowicz2012social, Szell2010, Szell2012social, Vsuvakov2013online}, like the mobile communication network. However, there are also counterexamples to the theory, one of which is the scientific collaboration network \cite{Ke2014, Ubaldi2021, Pan2012}. In this network, nodes represent authors, and an edge connects two authors if they co-authored at least one manuscript. The symmetric weight $w_{ij}$ equals the number of manuscripts co-authored by authors $i$ and $j$.

At first glance, the scientific collaboration network seems to defy Granovetter's theory, as neighbourhood overlap, on average, decreases with edge weight for the majority of edges. We have shown, however, that this supposed disagreement stems from improper definitions of weights and overlaps \cite{Fronczak2022}. Or, more specifically, from the fact that symmetric measures cannot properly describe the properties of this network.

\begin{figure}[t]
    \centering
    \includegraphics[width=\linewidth]{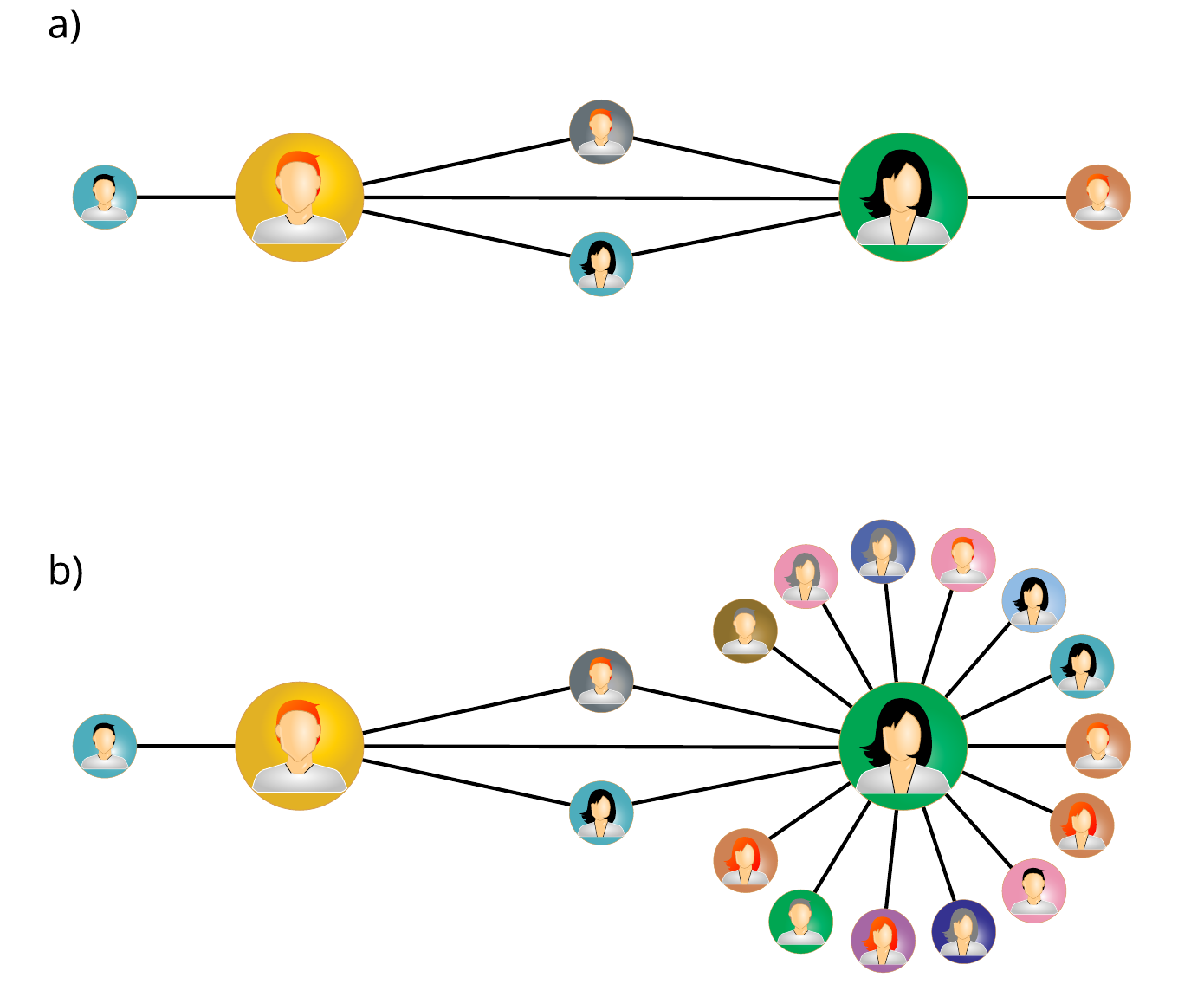}
    \caption{Examples of two connected nodes with a) similar and b) vastly different degrees.}\label{fig:overlaps}
\end{figure}
Fig.~\ref{fig:overlaps} illustrates the problem with symmetric measures. Panel a) shows two nodes, left $l$ and right $r$, with degrees $k_l = k_r = 4$. The nodes share two common neighbours. In this case, symmetric overlap equals
\begin{equation}
O_{lr} = O_{rl} = \frac{2}{(4 - 1) + (4 - 1) - 2} = \frac{2}{4} = \frac{1}{2}.
\end{equation}
Since the degrees of both nodes are identical, the two common neighbours constitute the same fraction of the neighbourhood of each node. In such a scenario,  symmetric overlap accurately reflects this observation from the standpoint of both nodes.
However, symmetric overlap fails when assessing nodes with vastly different degrees. In Fig.~\ref{fig:overlaps}b), the left node with $k_l = 4$ shares two common neighbours with the right node, whose degree is $k_r = 16$. Intuitively, the left node should assign much greater significance to the two common neighbours than the right node. Symmetric overlap cannot be used as a measure of this significance, as for both nodes it equals
\begin{equation}
O_{lr} = O_{rl} = \frac{2}{(4 - 1) + (16 - 1) - 2} = \frac{2}{16} = \frac{1}{8}.
\end{equation}
It is a low value, clearly skewed towards the node with the higher degree.

These two examples show that symmetric measures work properly when dealing with homogeneous networks, where we compare similar nodes - as was the case for many networks in which Granovetter's theory was proven to hold. Non-homogeneous networks, like the scientific collaboration network, which is scale-free \cite{Newman2010networks, Dorogovtsev2022nature} (there are often nodes with highly different degrees on both sides of an edge), require a different approach. The innate asymmetry of these networks suggests that one must use asymmetric measures instead of symmetric ones. In \cite{Fronczak2022}, we introduced the asymmetric overlap $Q$:
\begin{equation}\label{eq:Q}
Q_{ij} = \frac{n_{ij}}{k_i - 1},
\end{equation}
with
\begin{equation}
Q_{ij} \neq Q_{ji}.
\end{equation}
Returning to the example from Fig.~\ref{fig:overlaps}b), asymmetric overlap for the left node is 
\begin{equation}
Q_{lr} = \frac{2}{4 - 1} = \frac{2}{3},
\end{equation}
while for the right node, we have
\begin{equation}
Q_{rl} = \frac{2}{16 - 1} = \frac{2}{15}.
\end{equation}
These two values of overlap properly convey the importance of the shared neighbourhood from the perspective of each node separately. Asymmetric overlap reflects the asymmetric relationships of authors in the scientific collaboration network (and other non-homogeneous networks). What can be a large fraction of collaborators from the perspective of one author, can be a negligible fraction from the perspective of another author.

Symmetric definitions of weights suffer from similar issues in non-homogeneous networks. Symmetric weight in the scientific collaboration network usually equals the number of collaborations (co-authored articles) between two authors. However, the importance of a single collaboration depends on the total number of collaborations. If someone wrote only one paper in collaboration with an author who published tens or hundreds of manuscripts, then intuitively, the strength of that tie (the weight of the edge) should be greater from the perspective of the former author. Thus, in \cite{Fronczak2022}, we also introduced the asymmetric definition of weight $v$:
\begin{equation}\label{eq:v}
v_{ij} = \frac{w_{ij}}{m_i},
\end{equation}
where $w_{ij}$ is the symmetric weight, and $m_i$ is the number of papers published by the $i$-th author. For asymmetric weights, we also have
\begin{equation}
v_{ij} \neq v_{ji}.
\end{equation}

Using asymmetric overlaps and asymmetric weights, we showed that Granovetter's theory holds in the scientific collaboration network. We also postulated that these are natural and intuitive tools capable of properly describing scale-free networks, with application to other problems, eg. link prediction \cite{Orzechowski2023}. However, these asymmetric definitions introduce a certain non-obvious issue, especially when it comes to confirming Granovetter's theory.

The nature of this theory - or rather, the nature of the correlation between weights and overlaps postulated by Granovetter - is sociological. That is, the correlation must stem from actual social interactions between entities represented by nodes in the network. In contrast to that, measures defined in (\ref{eq:Q}) and (\ref{eq:v}) introduce structural correlations to the mix - correlations that result from the topology of the network. It is not unreasonable to assume that the number of papers published by an author ($m_i$) will relate in some way to the total number of collaborators ($k_i$) - intuitively, one can expect a positive correlation between these two variables. It raises the following questions: What are we really observing if we detect a correlation between asymmetric overlap and asymmetric weight? What is the source of that correlation? Are we truly confirming Granovetter's theory, or are we merely misinterpreting the effects of the network's topology? This paper aims to dispel these doubts using tools introduced in the next section.

\section{Methods}

Let us reiterate the problem mentioned in the Introduction and define it in a clearer and more tangible way. The main assumption behind Granovetter's theory is that weights in social networks are not assigned to edges randomly. Instead, they quantify the strength of interpersonal interactions and follow various patterns dictated by the nature of these interactions. One such pattern is that the strength of interactions should be directly tied to the overlap between social circles of nodes. The higher the overlap, the greater the strength of interaction. It is an intuitive and relatable conclusion - for example, ties within a family, which is a densely connected social circle, should be stronger than ties within a workplace.

Assuming that Granovetter's theory is correct, we could expect that in a network in which weights are assigned completely at random, the correlation between overlap and weight does not exist at all. By the same token, if we were to randomize weights in a network by shuffling them among the edges, such a procedure should also destroy the correlation between overlaps and weights. Unfortunately, while this is true for symmetric measures, asymmetric weights and overlaps still exhibit correlation even with randomised weights. The source of these correlations was mentioned in the previous section, and it is the structural correlation between $m_i$ and $k_i$, which are in the denominators of the asymmetric measures.

In fact, for reasons that will be explained in detail in the next section, we will use a definition of asymmetric weight different to the one introduced in \cite{Fronczak2022}. In this work, asymmetric weight will be defined as (cf. Eq.~\ref{eq:v})
\begin{equation}\label{eq:vv}
v_{ij} = \frac{w_{ij}}{s_i},
\end{equation}
where $s_i$ is the strength of the $i$-th node
\begin{equation}
s_{i} = \sum_j w_{ij}.
\end{equation}
Here, the structural correlations are even clearer. Since $s_i \propto k_i$ \cite{Barrat2004architecture} (for example, if we assume that all weights $w_{ij} = 1$, then $s_i$ is equal to $k_i$), 
\begin{equation}
v_{ij} \propto \frac{1}{s_i},
\end{equation}
and
\begin{equation}
Q_{ij} \propto \frac{1}{k_i},
\end{equation}
we must have
\begin{equation}
Q_{ij} \propto v_{ij}.
\end{equation}
The existence of the correlation between $Q_{ij}$ and $v_{ij}$ is largely independent of the distribution of symmetric weights $w_{ij}$ in the graph. That is, if we were to shuffle existing weights between edges or assign completely new weights to edges according to some probability distribution, this structural correlation would still be present.

The challenge is, then, to decouple the structural correlations from Granovetter-like social correlations while keeping the asymmetric definitions of strengths and overlaps. Thankfully, this is hardly a new kind of problem, and there are tools capable of dealing with similar issues. More specifically, we are going to employ so-called correlation profiles \cite{Maslov2002, Maslov2004}, which were used before to study mixing patterns in complex networks (correlations between node degrees at the ends of the same edge) \cite{Newman2002mixing, Newman2003mixing, Litvak2013uncovering}.

The idea behind correlation profiles is simple but powerful. One needs to compare the properties of the actual network with the properties of its randomised realisations (the null model). If the null model is chosen correctly, then the difference between random realisations and the actual network should result not from structural correlations but, in our case, from sociological processes (which are not present in the null model) that govern the assignment of weights to edges.

Correlation profiles are constructed using two simple ratios. If we want to study some pattern $p$ observed in a network, then we have to compare the number $N(p)$ of occurrences of that pattern in the actual network with the average number $\langle{N}_r(p)\rangle$ of occurrences of the same pattern in randomised realisations of the network. Using these two numbers, we can define the ratio
\begin{equation}\label{eq:R}
R(p) = \frac{N(p)}{\langle{N}_r(p)\rangle}.
\end{equation}
If $R(p)$ is close to 1, then there is no significant difference between the null model and the actual network. It follows that pattern $p$ is associated with properties captured in the null model. On the other hand, if $R(p)$ is higher or lower than 1, then there are mechanisms in the actual network that are responsible for the creation (or dissolution) of pattern $p$ that are not present in the null model.

The second ratio - Z-score -  is defined as 
\begin{equation}
Z(p) = \frac{N(p) - \langle{N}_r(p)\rangle}{\Delta N_r(p)}, 
\end{equation}
where $\Delta N_r(p)$ is the standard deviation of $N_r(p)$ in the randomised realisations of the network. This ratio determines the statistical significance of $R(p)$.

In most cases, correlation profiles are represented as two-dimensional images. To give a more concrete example, if we want to study the relation between overlap $Q$ and weight $v$, we divide the $Q - v$ plane into two-dimensional bins of equal size on a log-log scale (we use a logarithmic scale because $Q$ and $v$ values span multiple decades). Patterns $p$ correspond to pairs $(v, Q)$ (each edge in the network introduces two such pairs) falling into corresponding bins.

We count the number of points $N(p_i)$ that fall into the $i$-th bin in the actual network (here, $p_i$ denotes a pattern corresponding to a point falling into the $i$-th bin). Next, we create many random realisations of the network by shuffling symmetric weights and average over these realisations the number of points $N_r(p_i)$ that fall into the corresponding bin. Dividing these two numbers gives us the ratio $R(p_i)$ for the $i$-th bin. We repeat this procedure for each bin (using the same random realisations), which gives us the full correlation profile. Z-scores are calculated in the same way. 

\begin{figure}[t]
    \centering
    \includegraphics[width=\linewidth]{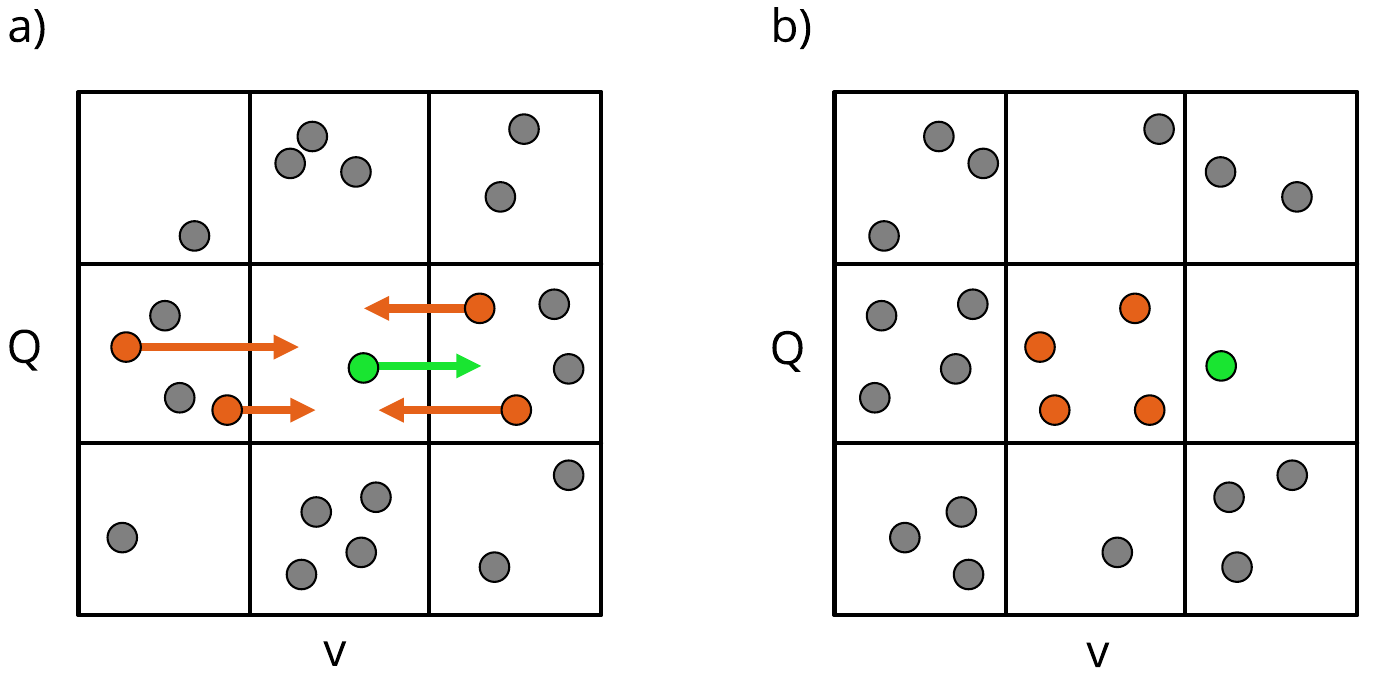}
    \caption{An example illustrating the creation of correlation profiles - points in bins a) before weight shuffling and b) after shuffling.}\label{fig:shuffling}
\end{figure}
An illustration of $R(p_i)$ calculation can be found in Fig.~\ref{fig:shuffling}. In this example, we concentrate on the middle bin. When weights are shuffled among edges during the creation of randomised graph instances (the null model), points on the $Q - v$ diagrams change their positions. However, they only move along the $v$ axis. The overlaps, which are independent of weights, do not change. Since the network's topology is fully retained during weight shuffling, the null model leaves the structural correlations intact. In Fig.~\ref{fig:shuffling}, points that will move into the middle bin after shuffling are orange, while the point that will move out of the middle bin is green. Arrows indicate where each of the relevant points will end up after shuffling. Thus, for the middle bin, we have
\begin{equation}\label{eq:R}
R(\text{middle bin}) = \frac{1}{4}.
\end{equation}
This value of $R$ suggests that the processes responsible for the distribution of weights in the actual network remove points from the middle bin when compared with a random instance of the network, possibly prioritizing other bins in the diagram. While it is an oversimplification (especially since we used only one randomised network instance instead of an entire ensemble, as required by the definition in Eq.~\ref{eq:R}), this example demonstrates the main idea behind the correlation diagrams. The non-structural correlations can be singled out by comparing the positions of points on the $Q - v$ diagrams corresponding to the actual network and its randomised instances.

\section{Datasets}
\begin{table}[]
\centering
\label{tab:datasets}
\begin{tabular}{|l|l|l|l|}
\hline
Dataset                            & Nodes   & Edges   & Type \\ \hline\hline
Twitter                            & 510136  & 5210278  & directed \\ \hline
DBLP                               & 2851120 & 24965776 & bipartite \\ \hline
Actor Movies                       & 374511  & 30029678 & bipartite \\ \hline
Record Labels                      & 11078   & 117798   & bipartite \\ \hline
Marvel & 6403    & 343176   & bipartite \\ \hline
Flights                            & 1292    & 24925    & directed \\ \hline
Metabolic Network                  & 3160    & 29210    & bipartite \\ \hline
Caenorhabditis Elegans    & 297     & 4296     & directed \\ \hline
\end{tabular}
\caption{Sizes of largest connected components in the datasets.}
\end{table}

In \cite{Fronczak2022}, we studied the validity of Granovetter's theory only in the scientific collaboration network. In this work, wanting to test both the theory itself and the applicability of correlation profiles on a variety of different networks, we used 8 datasets in total:
\begin{itemize}
    \item Twitter (source: \footnote{https://figshare.com/articles/dataset/Emergence\_and\_evolution\_of\_social\_networks\_through\_exploration\_of\_the\_Adjacent\_Possible/13308428}) - the network of Twitter mentions \cite{Ubaldi2021}. Nodes represent Twitter users and weights correspond to the number of mentions. 
    \item DBLP (source: \footnote{https://www.aminer.org/citation}) - the scientific collaboration network (version 12). It contains metadata about scientific articles \cite{Tang2008}, including lists of authors and references. Nodes represent authors; two authors are connected if they co-authored at least one paper. Symmetric weight is equal to the total number of papers co-authored by two authors.
    \item Actor Movies (source: \footnote{http://konect.cc/networks/actor-movie/}) - nodes represent actors, two actors are connected if they appeared in the same film. Symmetric weight is equal to the number of films in which actors worked together. 
    \item Record Labels (source: \footnote{http://konect.cc/networks/dbpedia-recordlabel/}) - nodes represent music artists, two artists are connected if they performed under the same record labels. Symmetric weight is equal to the number of record labels under which artists worked together. 
    \item The Marvel Universe Social Network (source: \footnote{https://www.kaggle.com/datasets/csanhueza/the-marvel-universe-social-network/}) - nodes represent heroes, two heroes are connected if they appeared in the same comic \cite{Alberich2002marvel}. Symmetric weight is equal to the number of comics in which heroes appeared together.    
    \item Flights - network of passenger flights. Nodes represent airports, and weights correspond to the volume of traffic (number of passengers) between airports. This database is commercial and is not publicly available.    
    \item Metabolic Network (source: \footnote{https://www.ebi.ac.uk/biomodels/MODEL6399676120}) - where nodes represent reactants, connected by an edge when they take part in the same reaction \cite{Li2010}. Symmetric weight equals the number of reactions sharing two given reactants.
    \item Caenorhabditis Elegans (source: \footnote{http://konect.cc/networks/dimacs10-celegansneural/}) - the neural network of \textit{Caenorhabditis elegans} \cite{Watts1998}. Nodes represent neurons, and an edge links two neurons if a synapse or gap junction connects them. Weights correspond to the total number of connections between neurons.
\end{itemize}
Not all of these networks are social networks, and some are artificial social networks. However, they all exhibit a Granovetter-like relationship between overlaps and weights. Table~\ref{tab:datasets} contains information about the sizes of the largest connected components in the networks - our analysis was constrained to these components.

Some of the networks we used can be represented as bipartite graphs (e.g. DBLP, Actor Movies - virtually all collaboration networks can be stored in this form) and recovered via appropriate projections \cite{Newman2003different, Zhou2007bipartite}. These networks are undirected and have a well-defined notion of symmetric weight. One can also easily use (\ref{eq:v}) to define asymmetric weights in such networks, with $m_i$ equal to the degree of node $i$ in the bipartite representation of a graph (which corresponds to the total number of collaborations for a given node - e.g. movies or scientific manuscripts). On the other hand, networks like Twitter or Flights are inherently directed, cannot be expressed as bipartite graphs and, consequently, Eq.~(\ref{eq:v}) cannot be applied in a meaningful way.

In order to standardise our approach to the networks under study and overcome problems associated with Eq.~(\ref{eq:v}), we decided to symmetrise all directed networks and assumed that symmetric weight in their undirected equivalent is equal to the average of weights in both directions:
\begin{equation}\label{eq:V}
    w_{ij} = \frac{V_{ij} + V_{ji}}{2},
\end{equation}
where $V_{ij}$ and $V_{ji}$ are weights of directed edges. At the same time, we abandoned the definition of asymmetric weight introduced in \cite{Fronczak2022}, and settled on definition (\ref{eq:vv}) instead (where asymmetry is achieved by normalising symmetric weight - that is by dividing it by the strength of a node). While it may seem as counter intuitive - directed networks are converted to undirected ones using Eq.~\ref{eq:V}, only to be converted again to directed networks using Eq.~\ref{eq:vv} - this approach allows us to treat all networks, both directed and undirected ones, in the same way and to compare results.

\section{Results}
\begin{figure*}[t]
    \centering
    \includegraphics[width=\linewidth]{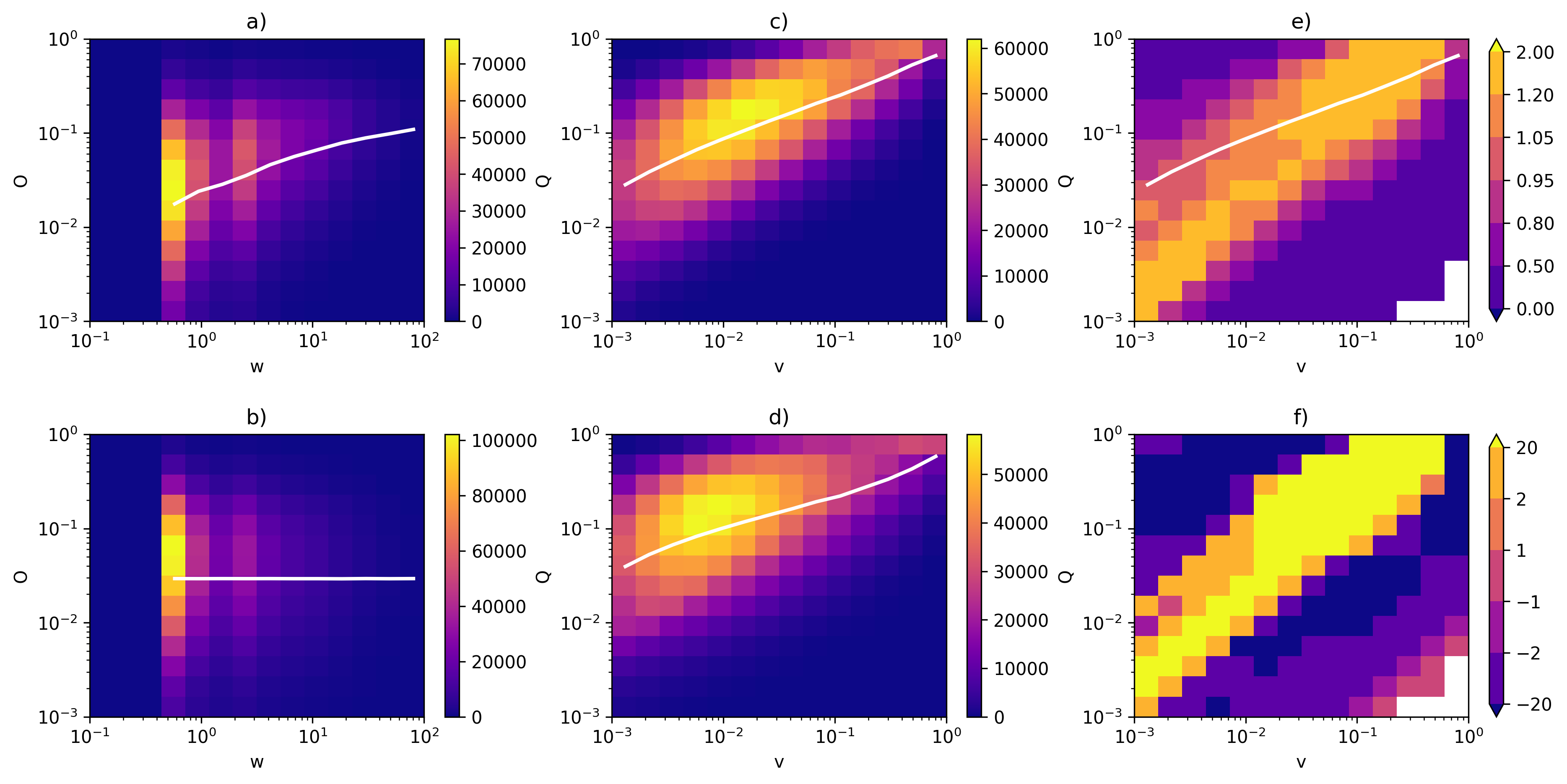}
    \caption{Correlation profiles for Twitter. a) Heatmap for the actual network - symmetric weights. b) Heatmap for the null model (randomised network) - symmetric weights. c) Heatmap for the actual network - asymmetric weights. d) Heatmap for the null model (randomised network) - asymmetric weights. e) Correlation profile (R). f) Z-score (Z). The white lines in a) and b) correspond to the average $O$ as a function of $w$, on c) and d) - the average $Q$ as a function of $v$. The line in panel e) is the same as in c).}\label{fig:twitter}
\end{figure*}
\begin{figure*}[t]
    \centering
    \includegraphics[width=\linewidth]{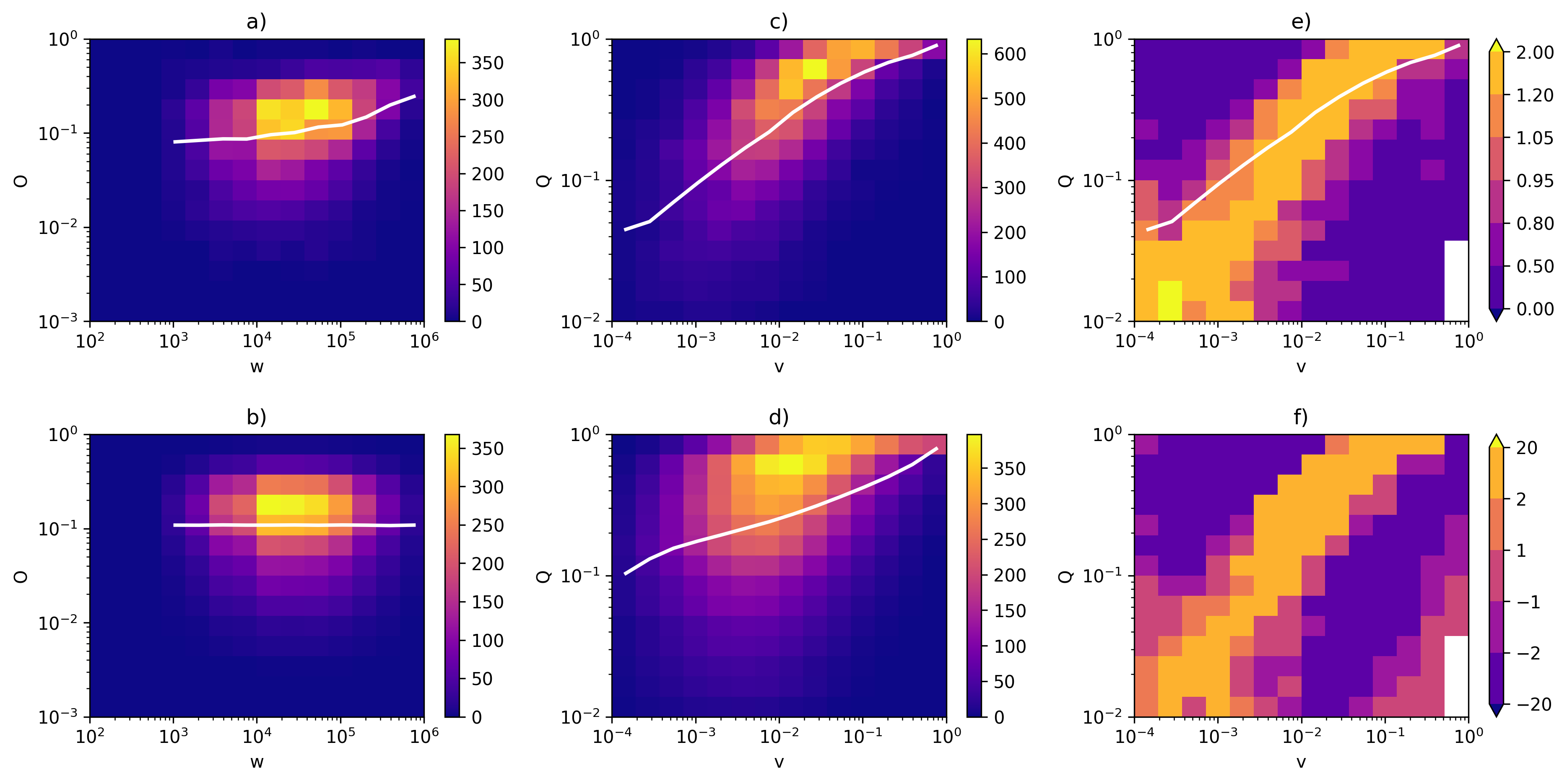}    
    \caption{Correlation profiles for the network of flights. a) Heatmap for the actual network - symmetric weights. b) Heatmap for the null model (randomised network) - symmetric weights. c) Heatmap for the actual network - asymmetric weights. d) Heatmap for the null model (randomised network) - asymmetric weights. e) Correlation profile (R). f) Z-score (Z). The white lines in a) and b) correspond to the average $O$ as a function of $w$, on c) and d) - the average $Q$ as a function of $v$. The line in panel e) is the same as in c).}\label{fig:flights}
\end{figure*}
\begin{figure*}[t]
    \centering
    \includegraphics[width=\linewidth]{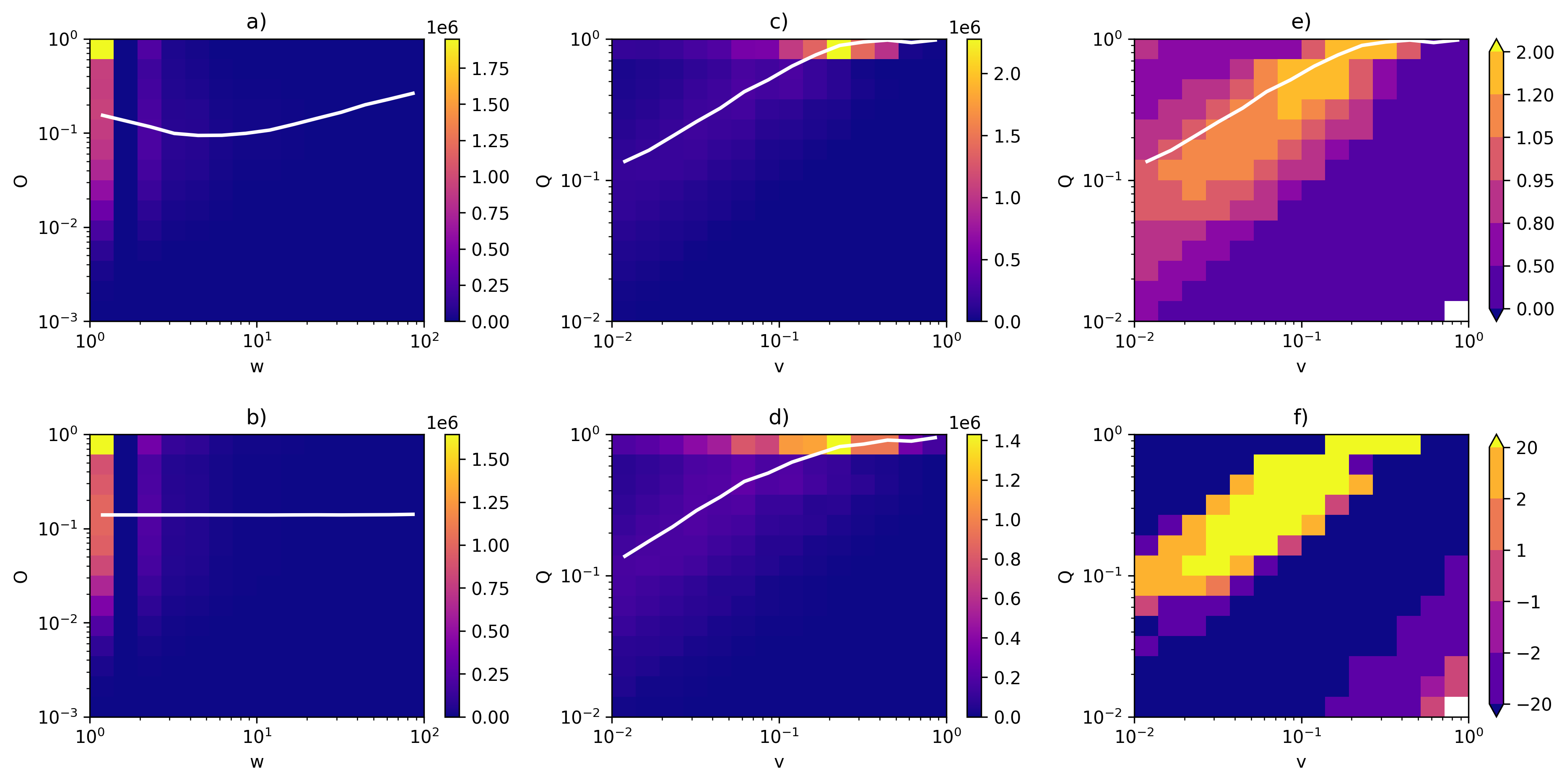}
    \caption{Correlation profiles for DBLP. a) Heatmap for the actual network - symmetric weights. b) Heatmap for the null model (randomised network) - symmetric weights. c) Heatmap for the actual network - asymmetric weights. d) Heatmap for the null model (randomised network) - asymmetric weights. e) Correlation profile (R). f) Z-score (Z). The white lines in a) and b) correspond to the average $O$ as a function of $w$, on c) and d) - the average $Q$ as a function of $v$. The line in panel e) is the same as in c).}\label{fig:dblp}
\end{figure*}
Correlational profiles for Twitter, a real social network, are shown in Fig.~\ref{fig:twitter}. Panels a) and b) contain, for comparison with their asymmetric counterparts, heatmaps of the symmetric overlap $O$ as a function of symmetric weight $w$ for the actual network and the null model. It is worth noting that, in many cases, symmetric weights are integers, and edges are often characterized by the same weight values. This makes edges indistinguishable from one another, which is a problem associated with using symmetric weights. Asymmetric weights are free of this issue, which is their additional benefit.

Panel c) contains heatmaps for the asymmetric overlap $Q$ as a function of asymmetric weight $v$. A clear, Granovetter-like relation is visible - overlap increases with weight. However, almost the same relation is present in panel d), which contains the equivalent heatmap for the null model (the same network with shuffled edges). These two panels show the root of the issue with the asymmetric definitions of weights and overlaps. Granovetter's theory dictates how weights should be distributed in a graph. If the theory is correct, then we should reasonably expect that there is no correlation between $Q$ and $v$ in the null model - the shuffling procedure should destroy any deliberate (from the perspective of Granovetter's theory) placement of weights. Unfortunately, such a correlation is also present in the null model due to the network's topology. Moreover, at first glance, the relation between $Q$ and $v$ seems to be very similar in the actual network and the null model.

This is where the correlation profiles come into play. By comparing panels c) and d) - that is, by dividing counts in bins in c) by counts in corresponding bins in d), which creates the correlation profile $R$, Eq.~\ref{eq:R} - we can easily find the differences between the null model and the real network. Panel e) shows such a profile. We can also see a Granovetter-like relation visible there - linear (on a log-log scale) clusters of bins such that more edges fall into these clusters in the actual networks than in the null model. It strongly suggests that Granovetter's theory is indeed correct and that sociological processes that govern the distribution of weights in real networks result in higher weights assigned to edges with higher values of overlaps. These results are statistically significant, which is confirmed by Z-scores in panels f).

We calculated correlation profiles and Z-scores for all networks presented in the previous section. More examples can be found in Fig.~\ref{fig:flights} and Fig.~\ref{fig:dblp}, which show profiles for the network of flights and DBLP. Note that in the case of DBLP, the average symmetric overlap is a decreasing function of symmetric weight for the majority of samples - it is precisely this behaviour that necessitates the introduction of asymmetric measures. Results for asymmetric measures presented in both figures are qualitatively equivalent to the ones in Fig.~\ref{fig:twitter}. Once again, we can see a correlation between $Q$ and $v$ in both the actual network and the null model. A Granovetter-like relation is prominent in panel e), suggesting that the processes responsible for the distribution of weights in this network prefer to assign higher weight values to edges characterised by higher overlap values. This observation holds true for all the networks examined in our study.

\begin{figure}
    \centering
    \includegraphics[width=\linewidth]{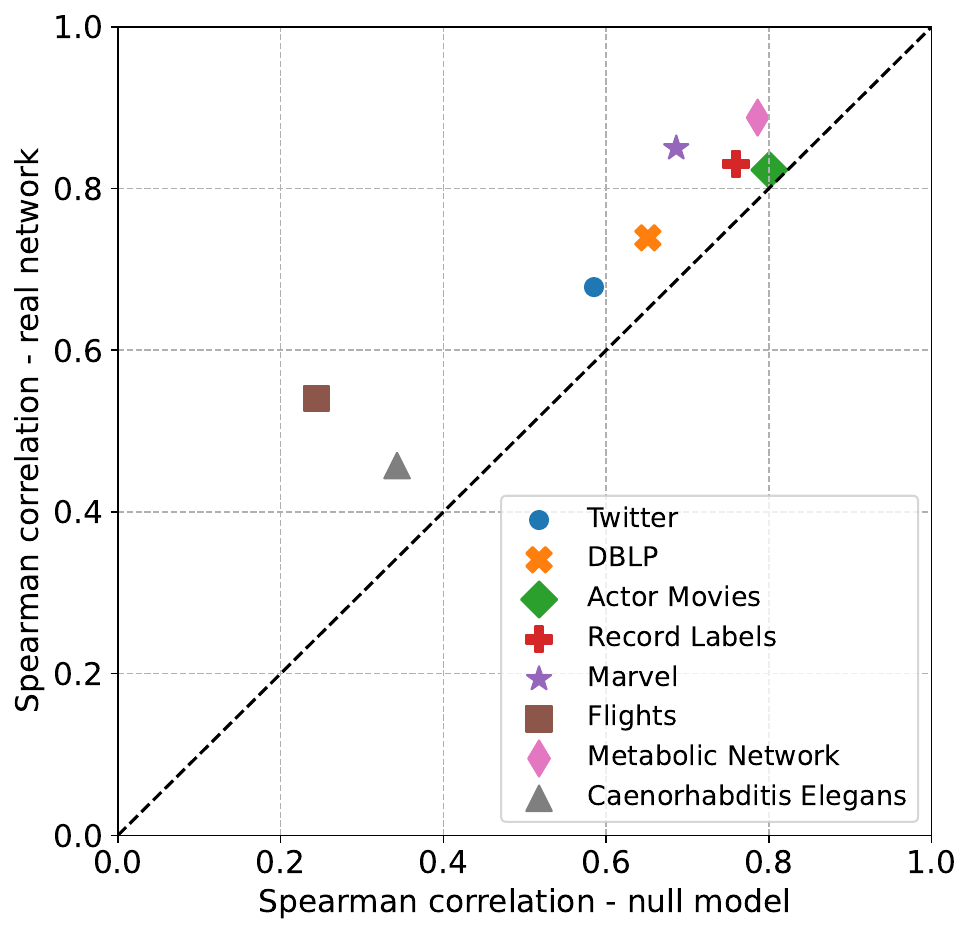}
    \caption{Spearman correlation between asymmetric overlaps and asymmetric weights in the real network as a function of the corresponding correlation in the null model.}\label{fig:corspe}
\end{figure}
There is another way to test Granovetter's theory - it is possible to calculate the correlation between overlaps and weights for the null model and the actual network. If Granovetter's theory is correct, then correlations in the real network should be stronger than in the null model. Fig.~\ref{fig:corspe} shows these correlations for all networks we studied. Considering the non-linearity of data, we decided to use the Spearman correlation and calculate it for logarithms of weights and overlaps. As can be seen, in all cases, there is a stronger positive correlation between weights and overlaps in the actual network, which supports Granovetter's theory.

\section{Summary and concluding remarks}
Due to the asymmetric nature of many human interactions (or, more generally, any interactions), symmetric measures cannot be universally used to describe social networks \cite{Mattie2018, Fronczak2022}. As we have shown, asymmetry is required in order to deal with such networks properly. For example, asymmetric measures can be used to confirm Granovetter's theory in the network of scientific collaborations, which was considered a counterexample to said theory. However, asymmetric measures - depending on their definitions - are not easy to interpret and require careful and deliberate handling.

In the case of the asymmetric overlap $Q$ and asymmetric weight $v$, as defined in Eqs.~(\ref{eq:Q}) and (\ref{eq:vv}), the problem with interpretation stems from the superfluous correlations introduced by the definitions of these measures. In fact, there are two layers of correlation that one needs to be wary of when analysing the relationship between $Q$ and $v$. The first layer is purely structural, induced by the network's topology. The strength of a node $s$ (the sum of weights over edges connecting the node to its neighbours) is correlated with the node's degree, resulting in a correlation between $Q$ and $v$. The second layer of correlations, the one we are truly interested in when confirming Granovetter's theory, is tied to the sociological processes that govern the distribution of weight between edges in the network. We assume that higher weight values will be assigned to edges with higher overlap values, which is not obvious, unlike the previous correlation. The problem is that correlations from both sources overlap, and a method that would allow us to differentiate between them is needed.

In this paper, we have shown that correlation profiles can be used to achieve this goal. The idea behind them is simple but effective - by randomising weights in a graph (shuffling them), we destroy the second kind of correlations, leaving only the structural correlations intact. Then, by comparing weights in the actual graph with its randomisations, we can determine how exactly the sociological processes responsible for weight distribution in a given network assign weights to edges. Our analysis shows that in the network we studied, a clear Granovetter-like relationship is present in the correlation diagrams (see Fig.~\ref{fig:twitter}e for Twitter and Fig.~\ref{fig:flights}e for the network of flights). That is, higher weight values are assigned, on average, to edges with higher overlap values - to the point that a monotonic relation (in the average sense on a log-log plot) is visible in the diagrams. This result truly confirms Granovetter's theory.

Moreover, not only did we study social networks and artificial social networks, but we also calculated correlation profiles for different kinds of networks - for example, the neural network of Caenorhabditis elegans or the metabolic network. These networks also exhibit a Granovetter-like relation between overlaps and weights, which suggests that Granovetter's theory is a sociological manifestation of more general principles governing complex networks.

On the one hand, we believe that this result is intuitive, as one can generally expect that if two nodes share a large portion of their neighbourhoods, then the strength of the connection between these nodes will likely be high. On the other hand, we hypothesise that the recently popularised theory of hidden metric spaces \cite{Allard2017, Serrano2022, Boguna2021, Krioukov2010} can provide a more formal explanation of this phenomenon. According to this theory, the topology of some networks and the values of weights can be explained by the existence of metric spaces in which these networks can be embedded - the connections in the network are determined, roughly speaking, by the positions of nodes in the hidden space. Such a structured way of determining (or explaining the topology of) neighbourhoods of nodes and edge weights likely leads to a correlation between weights and overlaps. However, we must emphasise that it is still a hypothesis and a possible and interesting direction for future studies.

\section{Acknowledgments} 

This research was funded by the POB Research Centre Cybersecurity and Data Science of Warsaw University of Technology within the Excellence Initiative Program - Research University (ID-UB).

\bibliography{bibliography}

\begin{thebibliography}{41}%
\makeatletter
\providecommand \@ifxundefined [1]{%
 \@ifx{#1\undefined}
}%
\providecommand \@ifnum [1]{%
 \ifnum #1\expandafter \@firstoftwo
 \else \expandafter \@secondoftwo
 \fi
}%
\providecommand \@ifx [1]{%
 \ifx #1\expandafter \@firstoftwo
 \else \expandafter \@secondoftwo
 \fi
}%
\providecommand \natexlab [1]{#1}%
\providecommand \enquote  [1]{``#1''}%
\providecommand \bibnamefont  [1]{#1}%
\providecommand \bibfnamefont [1]{#1}%
\providecommand \citenamefont [1]{#1}%
\providecommand \href@noop [0]{\@secondoftwo}%
\providecommand \href [0]{\begingroup \@sanitize@url \@href}%
\providecommand \@href[1]{\@@startlink{#1}\@@href}%
\providecommand \@@href[1]{\endgroup#1\@@endlink}%
\providecommand \@sanitize@url [0]{\catcode `\\12\catcode `\$12\catcode
  `\&12\catcode `\#12\catcode `\^12\catcode `\_12\catcode `\%12\relax}%
\providecommand \@@startlink[1]{}%
\providecommand \@@endlink[0]{}%
\providecommand \url  [0]{\begingroup\@sanitize@url \@url }%
\providecommand \@url [1]{\endgroup\@href {#1}{\urlprefix }}%
\providecommand \urlprefix  [0]{URL }%
\providecommand \Eprint [0]{\href }%
\providecommand \doibase [0]{https://doi.org/}%
\providecommand \selectlanguage [0]{\@gobble}%
\providecommand \bibinfo  [0]{\@secondoftwo}%
\providecommand \bibfield  [0]{\@secondoftwo}%
\providecommand \translation [1]{[#1]}%
\providecommand \BibitemOpen [0]{}%
\providecommand \bibitemStop [0]{}%
\providecommand \bibitemNoStop [0]{.\EOS\space}%
\providecommand \EOS [0]{\spacefactor3000\relax}%
\providecommand \BibitemShut  [1]{\csname bibitem#1\endcsname}%
\let\auto@bib@innerbib\@empty
\bibitem [{\citenamefont {Granovetter}(1973)}]{granovetter1973strength}%
  \BibitemOpen
  \bibfield  {author} {\bibinfo {author} {\bibfnamefont {M.~S.}\ \bibnamefont
  {Granovetter}},\ }\bibfield  {title} {\bibinfo {title} {The strength of weak
  ties},\ }\href@noop {} {\bibfield  {journal} {\bibinfo  {journal} {Am. J.
  Sociol.}\ }\textbf {\bibinfo {volume} {78}},\ \bibinfo {pages} {1360}
  (\bibinfo {year} {1973})}\BibitemShut {NoStop}%
\bibitem [{\citenamefont {Granovetter}(2018)}]{granovetter2018getting}%
  \BibitemOpen
  \bibfield  {author} {\bibinfo {author} {\bibfnamefont {M.~S.}\ \bibnamefont
  {Granovetter}},\ }\href@noop {} {\emph {\bibinfo {title} {Getting a Job: A
  Study of Contacts and Careers}}}\ (\bibinfo  {publisher} {University of
  Chicago Press},\ \bibinfo {year} {2018})\BibitemShut {NoStop}%
\bibitem [{\citenamefont {Onnela}\ \emph {et~al.}(2007)\citenamefont {Onnela},
  \citenamefont {Saram{\"a}ki}, \citenamefont {Hyv{\"o}nen}, \citenamefont
  {Szab{\'o}}, \citenamefont {Lazer}, \citenamefont {Kaski}, \citenamefont
  {Kert{\'e}sz},\ and\ \citenamefont {Barab{\'a}si}}]{Onnela2007structure}%
  \BibitemOpen
  \bibfield  {author} {\bibinfo {author} {\bibfnamefont {J.-P.}\ \bibnamefont
  {Onnela}}, \bibinfo {author} {\bibfnamefont {J.}~\bibnamefont
  {Saram{\"a}ki}}, \bibinfo {author} {\bibfnamefont {J.}~\bibnamefont
  {Hyv{\"o}nen}}, \bibinfo {author} {\bibfnamefont {G.}~\bibnamefont
  {Szab{\'o}}}, \bibinfo {author} {\bibfnamefont {D.}~\bibnamefont {Lazer}},
  \bibinfo {author} {\bibfnamefont {K.}~\bibnamefont {Kaski}}, \bibinfo
  {author} {\bibfnamefont {J.}~\bibnamefont {Kert{\'e}sz}},\ and\ \bibinfo
  {author} {\bibfnamefont {A.-L.}\ \bibnamefont {Barab{\'a}si}},\ }\bibfield
  {title} {\bibinfo {title} {Structure and tie strengths in mobile
  communication networks},\ }\href@noop {} {\bibfield  {journal} {\bibinfo
  {journal} {PNAS}\ }\textbf {\bibinfo {volume} {104}},\ \bibinfo {pages}
  {7332} (\bibinfo {year} {2007})}\BibitemShut {NoStop}%
\bibitem [{\citenamefont {Easley}\ and\ \citenamefont
  {Kleinberg}(2010)}]{Easley2010}%
  \BibitemOpen
  \bibfield  {author} {\bibinfo {author} {\bibfnamefont {D.}~\bibnamefont
  {Easley}}\ and\ \bibinfo {author} {\bibfnamefont {J.}~\bibnamefont
  {Kleinberg}},\ }\href@noop {} {\emph {\bibinfo {title} {Networks, Crowds, and
  Markets: Reasoning about a Highly Connected World}}}\ (\bibinfo  {publisher}
  {Cambridge University Press},\ \bibinfo {year} {2010})\BibitemShut {NoStop}%
\bibitem [{\citenamefont {Eagle}\ \emph {et~al.}(2010)\citenamefont {Eagle},
  \citenamefont {Macy},\ and\ \citenamefont {Claxton}}]{Eagle2010network}%
  \BibitemOpen
  \bibfield  {author} {\bibinfo {author} {\bibfnamefont {N.}~\bibnamefont
  {Eagle}}, \bibinfo {author} {\bibfnamefont {M.}~\bibnamefont {Macy}},\ and\
  \bibinfo {author} {\bibfnamefont {R.}~\bibnamefont {Claxton}},\ }\bibfield
  {title} {\bibinfo {title} {Network diversity and economic development},\
  }\href@noop {} {\bibfield  {journal} {\bibinfo  {journal} {Science}\ }\textbf
  {\bibinfo {volume} {328}},\ \bibinfo {pages} {1029} (\bibinfo {year}
  {2010})}\BibitemShut {NoStop}%
\bibitem [{\citenamefont {Pajevic}\ and\ \citenamefont
  {Plenz}(2012)}]{Pajevic2012organization}%
  \BibitemOpen
  \bibfield  {author} {\bibinfo {author} {\bibfnamefont {S.}~\bibnamefont
  {Pajevic}}\ and\ \bibinfo {author} {\bibfnamefont {D.}~\bibnamefont
  {Plenz}},\ }\bibfield  {title} {\bibinfo {title} {The organization of strong
  links in complex networks},\ }\href@noop {} {\bibfield  {journal} {\bibinfo
  {journal} {Nat. Phys.}\ }\textbf {\bibinfo {volume} {8}},\ \bibinfo {pages}
  {429} (\bibinfo {year} {2012})}\BibitemShut {NoStop}%
\bibitem [{\citenamefont {Grabowicz}\ \emph {et~al.}(2012)\citenamefont
  {Grabowicz}, \citenamefont {Ramasco}, \citenamefont {Moro}, \citenamefont
  {Pujol},\ and\ \citenamefont {Eguiluz}}]{Grabowicz2012social}%
  \BibitemOpen
  \bibfield  {author} {\bibinfo {author} {\bibfnamefont {P.~A.}\ \bibnamefont
  {Grabowicz}}, \bibinfo {author} {\bibfnamefont {J.~J.}\ \bibnamefont
  {Ramasco}}, \bibinfo {author} {\bibfnamefont {E.}~\bibnamefont {Moro}},
  \bibinfo {author} {\bibfnamefont {J.~M.}\ \bibnamefont {Pujol}},\ and\
  \bibinfo {author} {\bibfnamefont {V.~M.}\ \bibnamefont {Eguiluz}},\
  }\bibfield  {title} {\bibinfo {title} {Social features of online networks:
  The strength of intermediary ties in online social media},\ }\href@noop {}
  {\bibfield  {journal} {\bibinfo  {journal} {PLoS One}\ }\textbf {\bibinfo
  {volume} {7}},\ \bibinfo {pages} {e29358} (\bibinfo {year}
  {2012})}\BibitemShut {NoStop}%
\bibitem [{\citenamefont {Szell}\ and\ \citenamefont
  {Thurner}(2010)}]{Szell2010}%
  \BibitemOpen
  \bibfield  {author} {\bibinfo {author} {\bibfnamefont {M.}~\bibnamefont
  {Szell}}\ and\ \bibinfo {author} {\bibfnamefont {S.}~\bibnamefont
  {Thurner}},\ }\bibfield  {title} {\bibinfo {title} {Measuring social dynamics
  in a massive multiplayer online game},\ }\href@noop {} {\bibfield  {journal}
  {\bibinfo  {journal} {Soc. Networks}\ }\textbf {\bibinfo {volume} {32}},\
  \bibinfo {pages} {313} (\bibinfo {year} {2010})}\BibitemShut {NoStop}%
\bibitem [{\citenamefont {Szell}\ and\ \citenamefont
  {Thurner}(2012)}]{Szell2012social}%
  \BibitemOpen
  \bibfield  {author} {\bibinfo {author} {\bibfnamefont {M.}~\bibnamefont
  {Szell}}\ and\ \bibinfo {author} {\bibfnamefont {S.}~\bibnamefont
  {Thurner}},\ }\bibfield  {title} {\bibinfo {title} {Social dynamics in a
  large-scale online game},\ }\href@noop {} {\bibfield  {journal} {\bibinfo
  {journal} {Adv. Complex Syst.}\ }\textbf {\bibinfo {volume} {15}},\ \bibinfo
  {pages} {1250064} (\bibinfo {year} {2012})}\BibitemShut {NoStop}%
\bibitem [{\citenamefont {{\v{S}}uvakov}\ \emph {et~al.}(2013)\citenamefont
  {{\v{S}}uvakov}, \citenamefont {Mitrovi{\'c}}, \citenamefont
  {Gligorijevi{\'c}},\ and\ \citenamefont {Tadi{\'c}}}]{Vsuvakov2013online}%
  \BibitemOpen
  \bibfield  {author} {\bibinfo {author} {\bibfnamefont {M.}~\bibnamefont
  {{\v{S}}uvakov}}, \bibinfo {author} {\bibfnamefont {M.}~\bibnamefont
  {Mitrovi{\'c}}}, \bibinfo {author} {\bibfnamefont {V.}~\bibnamefont
  {Gligorijevi{\'c}}},\ and\ \bibinfo {author} {\bibfnamefont {B.}~\bibnamefont
  {Tadi{\'c}}},\ }\bibfield  {title} {\bibinfo {title} {How the online social
  networks are used: Dialogues-based structure of myspace},\ }\href@noop {}
  {\bibfield  {journal} {\bibinfo  {journal} {J. R. Soc. Interface}\ }\textbf
  {\bibinfo {volume} {10}},\ \bibinfo {pages} {20120819} (\bibinfo {year}
  {2013})}\BibitemShut {NoStop}%
\bibitem [{\citenamefont {Ke}\ and\ \citenamefont {Ahn}(2014)}]{Ke2014}%
  \BibitemOpen
  \bibfield  {author} {\bibinfo {author} {\bibfnamefont {Q.}~\bibnamefont
  {Ke}}\ and\ \bibinfo {author} {\bibfnamefont {Y.-Y.}\ \bibnamefont {Ahn}},\
  }\bibfield  {title} {\bibinfo {title} {Tie strength distribution in
  scientific collaboration networks},\ }\href@noop {} {\bibfield  {journal}
  {\bibinfo  {journal} {Phys. Rev. E}\ }\textbf {\bibinfo {volume} {90}},\
  \bibinfo {pages} {032804} (\bibinfo {year} {2014})}\BibitemShut {NoStop}%
\bibitem [{\citenamefont {Ubaldi}\ \emph {et~al.}(2021)\citenamefont {Ubaldi},
  \citenamefont {Burioni}, \citenamefont {Loreto},\ and\ \citenamefont
  {Tria}}]{Ubaldi2021}%
  \BibitemOpen
  \bibfield  {author} {\bibinfo {author} {\bibfnamefont {E.}~\bibnamefont
  {Ubaldi}}, \bibinfo {author} {\bibfnamefont {R.}~\bibnamefont {Burioni}},
  \bibinfo {author} {\bibfnamefont {V.}~\bibnamefont {Loreto}},\ and\ \bibinfo
  {author} {\bibfnamefont {F.}~\bibnamefont {Tria}},\ }\bibfield  {title}
  {\bibinfo {title} {Emergence and evolution of social networks through
  exploration of the adjacent possible space},\ }\href
  {https://doi.org/10.1038/s42005-021-00527-1} {\bibfield  {journal} {\bibinfo
  {journal} {Commun. Phys.}\ }\textbf {\bibinfo {volume} {4}},\ \bibinfo
  {pages} {28} (\bibinfo {year} {2021})}\BibitemShut {NoStop}%
\bibitem [{\citenamefont {Pan}\ and\ \citenamefont
  {Saramäki}(2012)}]{Pan2012}%
  \BibitemOpen
  \bibfield  {author} {\bibinfo {author} {\bibfnamefont {R.~K.}\ \bibnamefont
  {Pan}}\ and\ \bibinfo {author} {\bibfnamefont {J.}~\bibnamefont
  {Saramäki}},\ }\bibfield  {title} {\bibinfo {title} {The strength of strong
  ties in scientific collaboration networks},\ }\href@noop {} {\bibfield
  {journal} {\bibinfo  {journal} {Europhys. Lett.}\ }\textbf {\bibinfo {volume}
  {97}},\ \bibinfo {pages} {18007} (\bibinfo {year} {2012})}\BibitemShut
  {NoStop}%
\bibitem [{\citenamefont {Fronczak}\ \emph {et~al.}(2022)\citenamefont
  {Fronczak}, \citenamefont {Mrowinski},\ and\ \citenamefont
  {Fronczak}}]{Fronczak2022}%
  \BibitemOpen
  \bibfield  {author} {\bibinfo {author} {\bibfnamefont {A.}~\bibnamefont
  {Fronczak}}, \bibinfo {author} {\bibfnamefont {M.~J.}\ \bibnamefont
  {Mrowinski}},\ and\ \bibinfo {author} {\bibfnamefont {P.}~\bibnamefont
  {Fronczak}},\ }\bibfield  {title} {\bibinfo {title} {Scientific success from
  the perspective of the strength of weak ties},\ }\href@noop {} {\bibfield
  {journal} {\bibinfo  {journal} {Sci. Rep.}\ }\textbf {\bibinfo {volume}
  {12}},\ \bibinfo {pages} {5074} (\bibinfo {year} {2022})}\BibitemShut
  {NoStop}%
\bibitem [{\citenamefont {Newman}(2010)}]{Newman2010networks}%
  \BibitemOpen
  \bibfield  {author} {\bibinfo {author} {\bibfnamefont {M.~E.~J.}\
  \bibnamefont {Newman}},\ }\href@noop {} {\emph {\bibinfo {title} {Networks:
  An Introduction}}}\ (\bibinfo  {publisher} {Oxford University Press},\
  \bibinfo {year} {2010})\BibitemShut {NoStop}%
\bibitem [{\citenamefont {Dorogovtsev}\ and\ \citenamefont
  {Mendes}(2022)}]{Dorogovtsev2022nature}%
  \BibitemOpen
  \bibfield  {author} {\bibinfo {author} {\bibfnamefont {S.}~\bibnamefont
  {Dorogovtsev}}\ and\ \bibinfo {author} {\bibfnamefont {J.}~\bibnamefont
  {Mendes}},\ }\href@noop {} {\emph {\bibinfo {title} {The Nature of Complex
  Networks}}}\ (\bibinfo  {publisher} {Oxford University Press},\ \bibinfo
  {year} {2022})\BibitemShut {NoStop}%
\bibitem [{\citenamefont {Orzechowski}\ \emph {et~al.}(2023)\citenamefont
  {Orzechowski}, \citenamefont {Mrowinski}, \citenamefont {Fronczak},\ and\
  \citenamefont {Fronczak}}]{Orzechowski2023}%
  \BibitemOpen
  \bibfield  {author} {\bibinfo {author} {\bibfnamefont {K.~P.}\ \bibnamefont
  {Orzechowski}}, \bibinfo {author} {\bibfnamefont {M.~J.}\ \bibnamefont
  {Mrowinski}}, \bibinfo {author} {\bibfnamefont {A.}~\bibnamefont
  {Fronczak}},\ and\ \bibinfo {author} {\bibfnamefont {P.}~\bibnamefont
  {Fronczak}},\ }\bibfield  {title} {\bibinfo {title} {Asymmetry of social
  interactions and its role in link predictability: The case of coauthorship
  networks},\ }\href@noop {} {\bibfield  {journal} {\bibinfo  {journal} {J.
  Informetr.}\ }\textbf {\bibinfo {volume} {17}},\ \bibinfo {pages} {101405}
  (\bibinfo {year} {2023})}\BibitemShut {NoStop}%
\bibitem [{\citenamefont {Barrat}\ \emph {et~al.}(2004)\citenamefont {Barrat},
  \citenamefont {Barthélemy}, \citenamefont {Pastor-Satorras},\ and\
  \citenamefont {Vespignani}}]{Barrat2004architecture}%
  \BibitemOpen
  \bibfield  {author} {\bibinfo {author} {\bibfnamefont {A.}~\bibnamefont
  {Barrat}}, \bibinfo {author} {\bibfnamefont {M.}~\bibnamefont {Barthélemy}},
  \bibinfo {author} {\bibfnamefont {R.}~\bibnamefont {Pastor-Satorras}},\ and\
  \bibinfo {author} {\bibfnamefont {A.}~\bibnamefont {Vespignani}},\ }\bibfield
   {title} {\bibinfo {title} {The architecture of complex weighted networks},\
  }\href@noop {} {\bibfield  {journal} {\bibinfo  {journal} {PNAS}\ }\textbf
  {\bibinfo {volume} {101}},\ \bibinfo {pages} {3747} (\bibinfo {year}
  {2004})}\BibitemShut {NoStop}%
\bibitem [{\citenamefont {Maslov}\ and\ \citenamefont
  {Sneppen}(2002)}]{Maslov2002}%
  \BibitemOpen
  \bibfield  {author} {\bibinfo {author} {\bibfnamefont {S.}~\bibnamefont
  {Maslov}}\ and\ \bibinfo {author} {\bibfnamefont {K.}~\bibnamefont
  {Sneppen}},\ }\bibfield  {title} {\bibinfo {title} {Specificity and stability
  in topology of protein networks},\ }\href@noop {} {\bibfield  {journal}
  {\bibinfo  {journal} {Science}\ }\textbf {\bibinfo {volume} {296}},\ \bibinfo
  {pages} {910} (\bibinfo {year} {2002})}\BibitemShut {NoStop}%
\bibitem [{\citenamefont {Maslov}\ \emph {et~al.}(2004)\citenamefont {Maslov},
  \citenamefont {Sneppen},\ and\ \citenamefont {Zaliznyak}}]{Maslov2004}%
  \BibitemOpen
  \bibfield  {author} {\bibinfo {author} {\bibfnamefont {S.}~\bibnamefont
  {Maslov}}, \bibinfo {author} {\bibfnamefont {K.}~\bibnamefont {Sneppen}},\
  and\ \bibinfo {author} {\bibfnamefont {A.}~\bibnamefont {Zaliznyak}},\
  }\bibfield  {title} {\bibinfo {title} {Detection of topological patterns in
  complex networks: Correlation profile of the internet},\ }\href@noop {}
  {\bibfield  {journal} {\bibinfo  {journal} {Physica A}\ }\textbf {\bibinfo
  {volume} {333}},\ \bibinfo {pages} {529} (\bibinfo {year}
  {2004})}\BibitemShut {NoStop}%
\bibitem [{\citenamefont {Newman}(2002)}]{Newman2002mixing}%
  \BibitemOpen
  \bibfield  {author} {\bibinfo {author} {\bibfnamefont {M.~E.~J.}\
  \bibnamefont {Newman}},\ }\bibfield  {title} {\bibinfo {title} {Assortative
  mixing in networks},\ }\href@noop {} {\bibfield  {journal} {\bibinfo
  {journal} {Phys. Rev. Lett.}\ }\textbf {\bibinfo {volume} {89}},\ \bibinfo
  {pages} {208701} (\bibinfo {year} {2002})}\BibitemShut {NoStop}%
\bibitem [{\citenamefont {Newman}(2003)}]{Newman2003mixing}%
  \BibitemOpen
  \bibfield  {author} {\bibinfo {author} {\bibfnamefont {M.~E.~J.}\
  \bibnamefont {Newman}},\ }\bibfield  {title} {\bibinfo {title} {Mixing
  patterns in networks},\ }\href@noop {} {\bibfield  {journal} {\bibinfo
  {journal} {Phys. Rev. E}\ }\textbf {\bibinfo {volume} {67}},\ \bibinfo
  {pages} {026126} (\bibinfo {year} {2003})}\BibitemShut {NoStop}%
\bibitem [{\citenamefont {Litvak}\ and\ \citenamefont {van~der
  Hofstad}(2013)}]{Litvak2013uncovering}%
  \BibitemOpen
  \bibfield  {author} {\bibinfo {author} {\bibfnamefont {N.}~\bibnamefont
  {Litvak}}\ and\ \bibinfo {author} {\bibfnamefont {R.}~\bibnamefont {van~der
  Hofstad}},\ }\bibfield  {title} {\bibinfo {title} {Uncovering
  disassortativity in large scale-free networks},\ }\href@noop {} {\bibfield
  {journal} {\bibinfo  {journal} {Phys. Rev. E}\ }\textbf {\bibinfo {volume}
  {87}},\ \bibinfo {pages} {022801} (\bibinfo {year} {2013})}\BibitemShut
  {NoStop}%
\bibitem [{Note1()}]{Note1}%
  \BibitemOpen
  \bibinfo {note} {Https://figshare.com/articles/dataset/Emergence\protect
  \_and\protect \_evolution\protect \_of\protect \_social\protect
  \_networks\protect \_through\protect \_exploration\protect \_of\protect
  \_the\protect \_Adjacent\protect \_Possible/13308428}\BibitemShut {NoStop}%
\bibitem [{Note2()}]{Note2}%
  \BibitemOpen
  \bibinfo {note} {Https://www.aminer.org/citation}\BibitemShut {NoStop}%
\bibitem [{\citenamefont {Tang}\ \emph {et~al.}(2008)\citenamefont {Tang},
  \citenamefont {Zhang}, \citenamefont {Yao}, \citenamefont {Li}, \citenamefont
  {Zhang},\ and\ \citenamefont {Su}}]{Tang2008}%
  \BibitemOpen
  \bibfield  {author} {\bibinfo {author} {\bibfnamefont {J.}~\bibnamefont
  {Tang}}, \bibinfo {author} {\bibfnamefont {J.}~\bibnamefont {Zhang}},
  \bibinfo {author} {\bibfnamefont {L.}~\bibnamefont {Yao}}, \bibinfo {author}
  {\bibfnamefont {J.}~\bibnamefont {Li}}, \bibinfo {author} {\bibfnamefont
  {L.}~\bibnamefont {Zhang}},\ and\ \bibinfo {author} {\bibfnamefont
  {Z.}~\bibnamefont {Su}},\ }\bibfield  {title} {\bibinfo {title} {Arnetminer:
  Extraction and mining of academic social networks},\ }in\ \href@noop {}
  {\emph {\bibinfo {booktitle} {Proceedings of the 14th ACM SIGKDD
  International Conference on Knowledge Discovery and Data Mining}}},\ \bibinfo
  {series and number} {KDD '08}\ (\bibinfo  {publisher} {Association for
  Computing Machinery},\ \bibinfo {address} {New York, NY, USA},\ \bibinfo
  {year} {2008})\ p.\ \bibinfo {pages} {990–998}\BibitemShut {NoStop}%
\bibitem [{Note3()}]{Note3}%
  \BibitemOpen
  \bibinfo {note} {Http://konect.cc/networks/actor-movie/}\BibitemShut
  {NoStop}%
\bibitem [{Note4()}]{Note4}%
  \BibitemOpen
  \bibinfo {note} {Http://konect.cc/networks/dbpedia-recordlabel/}\BibitemShut
  {NoStop}%
\bibitem [{Note5()}]{Note5}%
  \BibitemOpen
  \bibinfo {note}
  {Https://www.kaggle.com/datasets/csanhueza/the-marvel-universe-social-network/}\BibitemShut
  {NoStop}%
\bibitem [{\citenamefont {Alberich}\ \emph {et~al.}(2002)\citenamefont
  {Alberich}, \citenamefont {Miro-Julia},\ and\ \citenamefont
  {Rossello}}]{Alberich2002marvel}%
  \BibitemOpen
  \bibfield  {author} {\bibinfo {author} {\bibfnamefont {R.}~\bibnamefont
  {Alberich}}, \bibinfo {author} {\bibfnamefont {J.}~\bibnamefont
  {Miro-Julia}},\ and\ \bibinfo {author} {\bibfnamefont {F.}~\bibnamefont
  {Rossello}},\ }\href@noop {} {\bibinfo {title} {Marvel universe looks almost
  like a real social network}} (\bibinfo {year} {2002}),\ \Eprint
  {https://arxiv.org/abs/cond-mat/0202174} {arXiv:cond-mat/0202174
  [cond-mat.dis-nn]} \BibitemShut {NoStop}%
\bibitem [{Note6()}]{Note6}%
  \BibitemOpen
  \bibinfo {note} {Https://www.ebi.ac.uk/biomodels/MODEL6399676120}\BibitemShut
  {NoStop}%
\bibitem [{\citenamefont {Li}\ \emph {et~al.}(2010)\citenamefont {Li},
  \citenamefont {Donizelli}, \citenamefont {Rodriguez}, \citenamefont
  {Dharuri}, \citenamefont {Endler}, \citenamefont {Chelliah}, \citenamefont
  {Li}, \citenamefont {He}, \citenamefont {Henry}, \citenamefont {Stefan},
  \citenamefont {Snoep}, \citenamefont {Hucka}, \citenamefont {Le~Nov{\`e}re},\
  and\ \citenamefont {Laibe}}]{Li2010}%
  \BibitemOpen
  \bibfield  {author} {\bibinfo {author} {\bibfnamefont {C.}~\bibnamefont
  {Li}}, \bibinfo {author} {\bibfnamefont {M.}~\bibnamefont {Donizelli}},
  \bibinfo {author} {\bibfnamefont {N.}~\bibnamefont {Rodriguez}}, \bibinfo
  {author} {\bibfnamefont {H.}~\bibnamefont {Dharuri}}, \bibinfo {author}
  {\bibfnamefont {L.}~\bibnamefont {Endler}}, \bibinfo {author} {\bibfnamefont
  {V.}~\bibnamefont {Chelliah}}, \bibinfo {author} {\bibfnamefont
  {L.}~\bibnamefont {Li}}, \bibinfo {author} {\bibfnamefont {E.}~\bibnamefont
  {He}}, \bibinfo {author} {\bibfnamefont {A.}~\bibnamefont {Henry}}, \bibinfo
  {author} {\bibfnamefont {M.~I.}\ \bibnamefont {Stefan}}, \bibinfo {author}
  {\bibfnamefont {J.~L.}\ \bibnamefont {Snoep}}, \bibinfo {author}
  {\bibfnamefont {M.}~\bibnamefont {Hucka}}, \bibinfo {author} {\bibfnamefont
  {N.}~\bibnamefont {Le~Nov{\`e}re}},\ and\ \bibinfo {author} {\bibfnamefont
  {C.}~\bibnamefont {Laibe}},\ }\bibfield  {title} {\bibinfo {title} {Biomodels
  database: An enhanced, curated and annotated resource for published
  quantitative kinetic models},\ }\href
  {https://doi.org/10.1186/1752-0509-4-92} {\bibfield  {journal} {\bibinfo
  {journal} {BMC Syst. Biol.}\ }\textbf {\bibinfo {volume} {4}},\ \bibinfo
  {pages} {92} (\bibinfo {year} {2010})}\BibitemShut {NoStop}%
\bibitem [{Note7()}]{Note7}%
  \BibitemOpen
  \bibinfo {note}
  {Http://konect.cc/networks/dimacs10-celegansneural/}\BibitemShut {NoStop}%
\bibitem [{\citenamefont {Watts}\ and\ \citenamefont
  {Strogatz}(1998)}]{Watts1998}%
  \BibitemOpen
  \bibfield  {author} {\bibinfo {author} {\bibfnamefont {D.~J.}\ \bibnamefont
  {Watts}}\ and\ \bibinfo {author} {\bibfnamefont {S.~H.}\ \bibnamefont
  {Strogatz}},\ }\bibfield  {title} {\bibinfo {title} {Collective dynamics of
  `small-world’ networks},\ }\href {https://doi.org/10.1038/30918} {\bibfield
   {journal} {\bibinfo  {journal} {Nature}\ }\textbf {\bibinfo {volume}
  {393}},\ \bibinfo {pages} {440} (\bibinfo {year} {1998})}\BibitemShut
  {NoStop}%
\bibitem [{\citenamefont {Newman}\ and\ \citenamefont
  {Park}(2003)}]{Newman2003different}%
  \BibitemOpen
  \bibfield  {author} {\bibinfo {author} {\bibfnamefont {M.~E.~J.}\
  \bibnamefont {Newman}}\ and\ \bibinfo {author} {\bibfnamefont
  {J.}~\bibnamefont {Park}},\ }\bibfield  {title} {\bibinfo {title} {Why social
  networks are different from other types of networks},\ }\href@noop {}
  {\bibfield  {journal} {\bibinfo  {journal} {Phys. Rev. E}\ }\textbf {\bibinfo
  {volume} {68}},\ \bibinfo {pages} {036122} (\bibinfo {year}
  {2003})}\BibitemShut {NoStop}%
\bibitem [{\citenamefont {Zhou}\ \emph {et~al.}(2007)\citenamefont {Zhou},
  \citenamefont {Ren}, \citenamefont {Medo},\ and\ \citenamefont
  {Zhang}}]{Zhou2007bipartite}%
  \BibitemOpen
  \bibfield  {author} {\bibinfo {author} {\bibfnamefont {T.}~\bibnamefont
  {Zhou}}, \bibinfo {author} {\bibfnamefont {J.}~\bibnamefont {Ren}}, \bibinfo
  {author} {\bibfnamefont {M.}~\bibnamefont {Medo}},\ and\ \bibinfo {author}
  {\bibfnamefont {Y.-C.}\ \bibnamefont {Zhang}},\ }\bibfield  {title} {\bibinfo
  {title} {Bipartite network projection and personal recommendation},\
  }\href@noop {} {\bibfield  {journal} {\bibinfo  {journal} {Phys. Rev. E}\
  }\textbf {\bibinfo {volume} {76}},\ \bibinfo {pages} {046115} (\bibinfo
  {year} {2007})}\BibitemShut {NoStop}%
\bibitem [{\citenamefont {Mattie}\ \emph {et~al.}(2018)\citenamefont {Mattie},
  \citenamefont {Eng{\o}-Monsen}, \citenamefont {Ling},\ and\ \citenamefont
  {Onnela}}]{Mattie2018}%
  \BibitemOpen
  \bibfield  {author} {\bibinfo {author} {\bibfnamefont {H.}~\bibnamefont
  {Mattie}}, \bibinfo {author} {\bibfnamefont {K.}~\bibnamefont
  {Eng{\o}-Monsen}}, \bibinfo {author} {\bibfnamefont {R.}~\bibnamefont
  {Ling}},\ and\ \bibinfo {author} {\bibfnamefont {J.-P.}\ \bibnamefont
  {Onnela}},\ }\bibfield  {title} {\bibinfo {title} {Understanding tie strength
  in social networks using a local ``bow tie’’ framework},\ }\href@noop {}
  {\bibfield  {journal} {\bibinfo  {journal} {Sci. Rep.}\ }\textbf {\bibinfo
  {volume} {8}},\ \bibinfo {pages} {9349} (\bibinfo {year} {2018})}\BibitemShut
  {NoStop}%
\bibitem [{\citenamefont {Allard}\ \emph {et~al.}(2017)\citenamefont {Allard},
  \citenamefont {Serrano}, \citenamefont {Garc{\'i}a-P{\'e}rez},\ and\
  \citenamefont {Bogu{\~{n}}{\'a}}}]{Allard2017}%
  \BibitemOpen
  \bibfield  {author} {\bibinfo {author} {\bibfnamefont {A.}~\bibnamefont
  {Allard}}, \bibinfo {author} {\bibfnamefont {M.~{\'A}.}\ \bibnamefont
  {Serrano}}, \bibinfo {author} {\bibfnamefont {G.}~\bibnamefont
  {Garc{\'i}a-P{\'e}rez}},\ and\ \bibinfo {author} {\bibfnamefont
  {M.}~\bibnamefont {Bogu{\~{n}}{\'a}}},\ }\bibfield  {title} {\bibinfo {title}
  {The geometric nature of weights in real complex networks},\ }\href@noop {}
  {\bibfield  {journal} {\bibinfo  {journal} {Nat. Commun.}\ }\textbf {\bibinfo
  {volume} {8}},\ \bibinfo {pages} {14103} (\bibinfo {year}
  {2017})}\BibitemShut {NoStop}%
\bibitem [{\citenamefont {Serrano}\ and\ \citenamefont
  {Bogu\~n\'a}(2022)}]{Serrano2022}%
  \BibitemOpen
  \bibfield  {author} {\bibinfo {author} {\bibfnamefont {M.~A.}\ \bibnamefont
  {Serrano}}\ and\ \bibinfo {author} {\bibfnamefont {M.}~\bibnamefont
  {Bogu\~n\'a}},\ }\href@noop {} {\emph {\bibinfo {title} {The Shortest Path to
  Network Geometry: A Practical Guide to Basic Models and Applications}}},\
  Elements in the Structure and Dynamics of Complex Networks\ (\bibinfo
  {publisher} {Cambridge University Press},\ \bibinfo {year}
  {2022})\BibitemShut {NoStop}%
\bibitem [{\citenamefont {Bogu\~n\'a}\ \emph {et~al.}(2021)\citenamefont
  {Bogu\~n\'a}, \citenamefont {Bonamassa}, \citenamefont {De~Domenico},
  \citenamefont {Havlin}, \citenamefont {Krioukov},\ and\ \citenamefont
  {Serrano}}]{Boguna2021}%
  \BibitemOpen
  \bibfield  {author} {\bibinfo {author} {\bibfnamefont {M.}~\bibnamefont
  {Bogu\~n\'a}}, \bibinfo {author} {\bibfnamefont {I.}~\bibnamefont
  {Bonamassa}}, \bibinfo {author} {\bibfnamefont {M.}~\bibnamefont
  {De~Domenico}}, \bibinfo {author} {\bibfnamefont {S.}~\bibnamefont {Havlin}},
  \bibinfo {author} {\bibfnamefont {D.}~\bibnamefont {Krioukov}},\ and\
  \bibinfo {author} {\bibfnamefont {M.~A.}\ \bibnamefont {Serrano}},\
  }\bibfield  {title} {\bibinfo {title} {Network geometry},\ }\href@noop {}
  {\bibfield  {journal} {\bibinfo  {journal} {Nat. Rev. Phys.}\ }\textbf
  {\bibinfo {volume} {3}},\ \bibinfo {pages} {114} (\bibinfo {year}
  {2021})}\BibitemShut {NoStop}%
\bibitem [{\citenamefont {Krioukov}\ \emph {et~al.}(2010)\citenamefont
  {Krioukov}, \citenamefont {Papadopoulos}, \citenamefont {Kitsak},
  \citenamefont {Vahdat},\ and\ \citenamefont {Bogu\~n\'a}}]{Krioukov2010}%
  \BibitemOpen
  \bibfield  {author} {\bibinfo {author} {\bibfnamefont {D.}~\bibnamefont
  {Krioukov}}, \bibinfo {author} {\bibfnamefont {F.}~\bibnamefont
  {Papadopoulos}}, \bibinfo {author} {\bibfnamefont {M.}~\bibnamefont
  {Kitsak}}, \bibinfo {author} {\bibfnamefont {A.}~\bibnamefont {Vahdat}},\
  and\ \bibinfo {author} {\bibfnamefont {M.}~\bibnamefont {Bogu\~n\'a}},\
  }\bibfield  {title} {\bibinfo {title} {Hyperbolic geometry of complex
  networks},\ }\href@noop {} {\bibfield  {journal} {\bibinfo  {journal} {Phys.
  Rev. E}\ }\textbf {\bibinfo {volume} {82}},\ \bibinfo {pages} {036106}
  (\bibinfo {year} {2010})}\BibitemShut {NoStop}%
\end{thebibliography}%

\end{document}